\documentclass[12pt, a4paper]{article}
\usepackage{graphicx}
\usepackage{chngpage}
\usepackage{xspace,ifthen,epsfig}
\usepackage{cite}
\usepackage{color}
\usepackage{caption}
\usepackage{fancybox}
\usepackage{float}
\usepackage{setspace}
\usepackage{subfigure}
\usepackage{longtable}
\usepackage{tabularx}
\usepackage{ltxtable}
\usepackage{times}
\usepackage[table]{xcolor}
\usepackage{url}
\usepackage{listings}
\usepackage{amsmath}
\usepackage{dsfont}
\usepackage[american]{babel}
\usepackage[utf8]{inputenc}
\usepackage{fancyhdr}
\usepackage{titling}
\usepackage{booktabs}
\usepackage{cancel}

\preauthor{\begin{center}
\normalsize \lineskip 0.5em%
\begin{tabular}[t]{c}}

\usepackage[%
   pdfauthor={Lehmann, Hubschle-Schneider, Sanders},
   pdftitle={Weighted Random Sampling on GPUs},
   pdfkeywords={Sampling, GPU}
]{hyperref}
\author{Hans-Peter Lehmann, Lorenz Hübschle-Schneider, Peter Sanders\\
    \texttt{\{h.lehmann, huebschle, sanders\}@kit.edu}\\
    Karlsruhe Institute of Technology}
\date{}

\newcounter{nalg}
\renewcommand{\thenalg}{\arabic{nalg}}
\DeclareCaptionLabelFormat{algocaption}{Algorithm \thenalg}

\lstnewenvironment{algorithm}[1][]
{   
    \refstepcounter{nalg}
    \captionsetup{labelformat=algocaption,labelsep=colon}
    \lstset{
        mathescape=true,
        frame=tb,
        numbers=none, %
        numberstyle=\small,
        basicstyle=\normalsize,
        columns=fullflexible,
        keywordstyle=\color{black}\bfseries,
        keywords={, return, function, if, then, else, foreach, while, kernel, unsigned, long, static, do,}
        xleftmargin=.04\textwidth,
        escapechar=^,
        #1 %
    }
}
{}

\newcommand{\evenindent}[2]{\ifodd #1 \else \hspace*{#2} \fi}

\newcommand{\rtx}[0]{RTX~2080\xspace}
\newcommand{\gtx}[0]{GTX~1650S\xspace}
\newcommand{\tesla}[0]{Tesla~V100\xspace}

\newcommand{\cc}[1]{\multicolumn{1}{c}{#1}}

\begin{document}
\unitlength1cm
\title{Weighted Random Sampling on GPUs}
\maketitle

\makebox{\centering
\parbox{0.9\textwidth}{
\paragraph{Abstract}
An alias table is a data structure that allows for efficiently drawing weighted random samples in constant time and can be constructed in linear time \cite{vose1991linear}. The PSA algorithm by Hübschle-Schneider and Sanders \cite{hubschle2019parallel} is able to construct alias tables in parallel on the CPU. In this report, we transfer the PSA algorithm to the GPU. Our construction algorithm achieves a speedup of 17 on a consumer GPU in comparison to the PSA method on a 16-core high-end desktop CPU. For sampling, we achieve an up to 24 times higher throughput. Both operations also require several times less energy than on the CPU. Adaptations helping to achieve this include changing memory access patterns to do coalesced access. Where this is not possible, we first copy data to the faster shared memory using coalesced access. We also enhance a generalization of binary search enabling to search for a range of items in parallel. Besides naive sampling, we also give improved batched sampling algorithms.
}}

\section{Introduction}
Weighted random sampling is the process of drawing items from a set $\{1, ..., N\}$, where each item has a specific weight $w_i \in \mathds{R}$. Denoting the total weight with $W=\sum_{1 \leq i \leq N}{w_i}$, each item is drawn with probability $\mathds{P}(i)=w_i/W$. In this report, we consider sampling with replacement, so the same item can be sampled multiple times. GPUs are becoming more important for high performance computing because of their fast memory and high degree of parallelism. Therefore, there is need for an efficient method to construct data structures for drawing weighted random samples on GPUs. A data structure that allows for efficiently sampling from a weighted random distribution in $\mathcal{O}(1)$ is the alias table, introduced by Walker \cite{walker1977efficient}.\\
Weighted random sampling has numerous applications, for example sampling recursion layers when generating R-MAT graphs \cite{hubschle2019linear}, sampling particle source positions in medical simulations \cite{wilderman2007method}, sampling ray directions in photorealistic rendering \cite{burke2004bidirectional}, and sampling word distributions in machine learning \cite{li2017saberlda}. Alias tables can also be used for interactive noise function generation \cite{galerne2012gabor}.\\
This report is based on and has text overlaps with the master's thesis of the first author \cite{lehmann2021alias}. The source code of the implementation is available on GitHub \cite{sourceGitHub}.

\section{Preliminaries}
\subsection{GPUs}
\paragraph{Basic Architecture.}
A GPU is highly symmetrical, consisting of multiple \emph{streaming multiprocessors}~(SMs). Each SM simultaneously executes multiple threads. The smallest level of parallelism, 32 threads, is called a \emph{warp}. All threads in a warp share their instruction pointer and inactive threads are masked out \cite{nvidia2018turingArchitecture}. Functions that are executed on the GPU are called \emph{kernels}. A kernel is executed on a grid of \emph{blocks}, each of which consists of a grid of threads that are scheduled to the same SM. Threads from the same block can synchronize and share memory, while threads from different blocks cannot cooperate directly \cite{nvidia2008gettingstarted}.

\paragraph{GPU Memory.}
The GPU has a large \emph{global memory} (also called \emph{device memory}). Additionally, each block can allocate \emph{shared memory} that is located directly on the SM and can be accessed much faster. Whenever the threads of a warp access global memory, the number of 32-byte transactions needed to fulfill the requests is minimized (\emph{coalescing}) \cite{nvidia2020bestpractices}. To leverage this performance improvement, special memory access patterns like \emph{interleaved addressing} need to be used. Moreover, the GPU's memory addresses are distributed over multiple physical memory modules called \emph{banks}. The banks can perform transactions in parallel but when multiple threads access the same bank in different rows, the operations need to be serialized \cite{nguyen2007gpu}.

\subsection{Alias Tables}
An alias table \cite{walker1977efficient} $T$ has $N$ rows, where $N$ is the number of items in the input set. Each row represents a bucket of equal share $W/N$ of the total weight. It has two columns, namely a weight $T^w_i \in \mathds{R}$ and an alias $T^a_i \in \{1, ..., N\}$. To sample, we draw a uniform random number $U \in (0, 1]$ and multiply it by $N$. The integer part $k=\lceil U \cdot N \rceil$ selects a row from the table. The fractional part is used to choose between item $k$ and its alias $T^a_k$ by checking if $\textit{frac}(U \cdot N) \cdot W/N < T^w_k$. Thus, alias tables allow for sampling an item in time $\mathcal{O}(1)$. It is possible to construct an alias table for every discrete distribution.

\begin{figure}[t]
    \centering
    \subfigure[Input weights]{\includegraphics[width=0.35\textwidth]{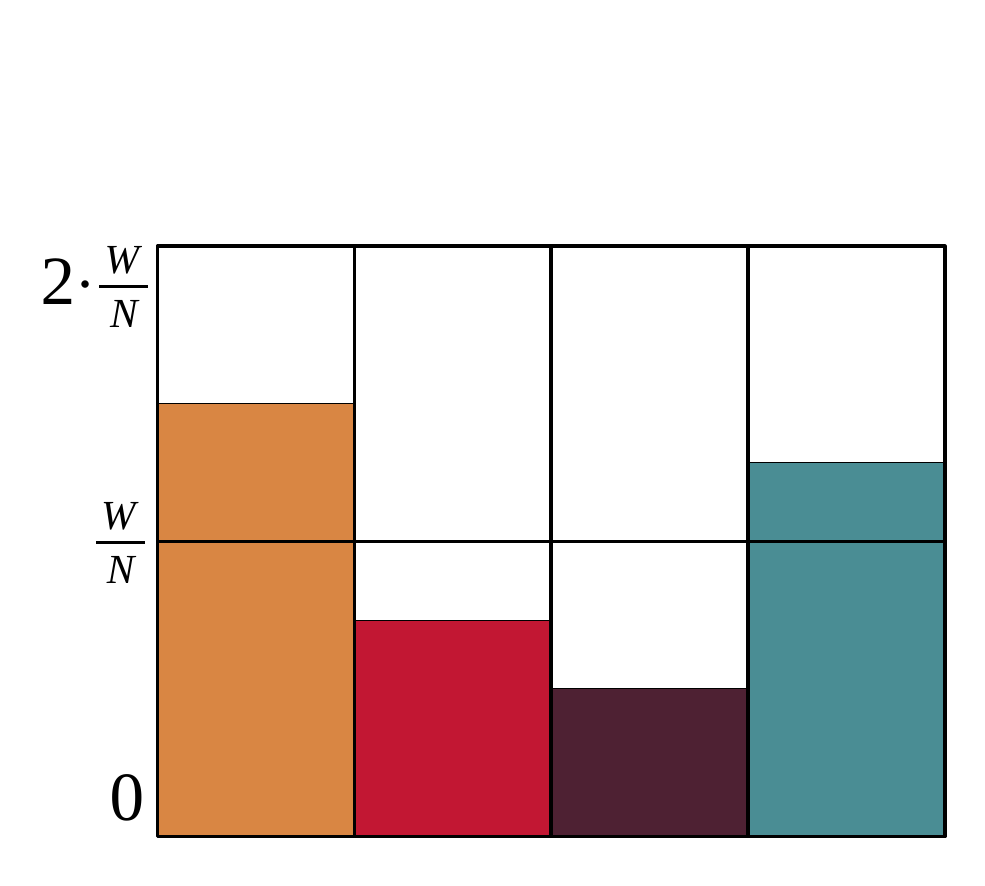}}
    \subfigure[Constructed table]{\includegraphics[width=0.35\textwidth]{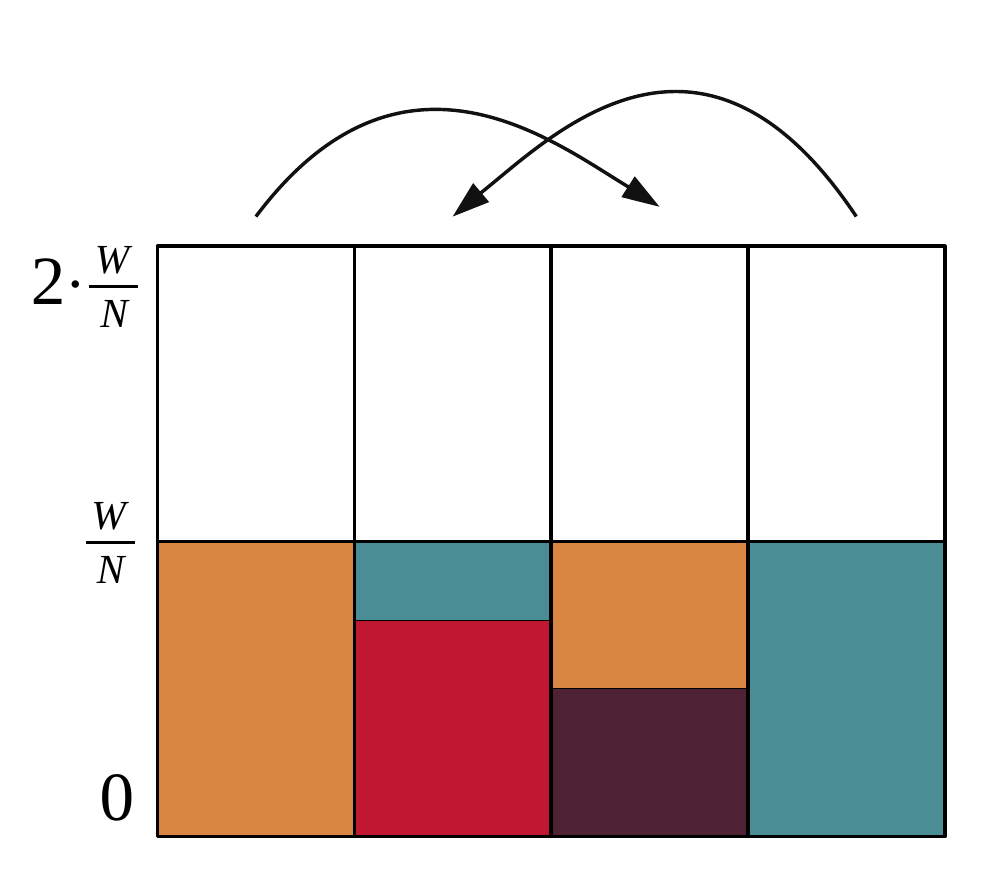}}
    \caption[Alias table]{Illustration of alias table construction. Buckets of items with weight smaller than the average are filled with excess weight of heavy items.}
    \label{fig:aliasTables}
\end{figure}

\paragraph{Sequential Construction.}
The idea of alias table construction is that \emph{heavy} items that are more likely to be sampled than a table row ($w_i > W/N$) give excess weight to the buckets of one or more \emph{light} items ($w_i \leq W/N$). This procedure is illustrated in Figure \ref{fig:aliasTables}. Vose \cite{vose1991linear} describes an $\mathcal{O}(N)$ alias table construction algorithm that explicitly maintains lists \texttt{l} and \texttt{h} of light and heavy items. While there are items available, the algorithm takes a heavy item $j \in \texttt{h}$. It then distributes the excess weight of that item by taking light items and filling their buckets. When a heavy item's weight drops below $W/N$, it is moved to the list of light items.

\paragraph{Parallel Construction.}
Hübschle-Schneider and Sanders' \cite{hubschle2019parallel} PSA method uses a two-step approach to parallel alias table construction. During the first step, splitting, the algorithm precomputes the state of Vose's construction at $s$ positions. These splits define sections that can later be worked on independently in parallel. The algorithm selects a number of light and heavy items in a way such that the number of items in each section is $N/s$ and the weights are balanced. Valid split positions are found by executing a binary search on the prefix sums of light and heavy items. Because the weight usually does not exactly fit into a section, the algorithm stores the remaining weight of the last heavy item as \emph{spill} to the next section. The result of the splitting step is a list of section boundaries and their respective spill values. The second step, packing, then constructs the actual alias table. In parallel, each processor iterates over the items of one of the sections and distributes weight from buckets of heavy items to buckets of light items. The PSA+ method \cite{hubschle2019parallel} is a semi-greedy variant that, instead of calculating prefix sums and splits for all items, builds the alias table in fixed-size sections until each section runs out of light or heavy items. PSA+ then only performs the PSA construction with the remaining items.

\section{Related Work}
Mohanty et al. \cite{mohanty2012efficient} implement only the alias table sampling step on the GPU and use it for Monte Carlo simulations. Binder and Keller~\cite{binder2019massively} introduce a \emph{monotonic} sampling algorithm for GPUs that is not based on alias tables and can sample in $\mathcal{O}(1)$ average case and in $\mathcal{O}(\log(N))$ worst case running time. A sampling algorithm is called \emph{monotonic} if a larger random number also generates a larger sample. This can be used to preserve the low discrepancy of quasi-random number generators. In this report, we do not consider the additional requirement of monotonicity and are rather interested in improving throughput.

\section{Construction}
Because our new method is based on PSA \cite{hubschle2019parallel}, we can now introduce our construction algorithm by explaining the splitting and packing steps individually.

\subsection{Split Method}
Let's denote the number of splits with the variable $s$. As a baseline, we transfer the original split algorithm of PSA \cite{hubschle2019parallel} directly to the GPU. Being based on binary search, the first iterations take the same branches and therefore read the same memory locations. This allows for coalescing but does not utilize the parallelism of the memory banks. We then introduce a new search operation that we call \emph{partial $p$-ary search} that makes use of both architectural properties. While we present it only in context of alias table construction, it can be used in other contexts, too.

\paragraph{Partial $p$-ary Search.}
For finding an item in a sorted list, Kaldewey et al. \cite{kaldewey2009parallel} evaluate $p$-ary search on GPUs. In contrast to binary search, $p$-ary search reduces the search range to $1/p$ in each iteration by looking at equally spaced pivots in parallel. The threads synchronize after each memory access and limit the search range to one of the sections. With plain $p$-ary search, all threads cooperate to search for one single item. Our new \emph{partial} $p$-ary search algorithm can be used to search for one item per thread. It makes use of the fact that the threads of a block often search for items that are close together in memory. The algorithm, to our knowledge, has not previously been described in the literature. The algorithm works in two phases. In the first phase, it executes $p$-ary search for all items of the block at once. In each iteration, instead of continuing the search on one section, partial $p$-ary search reduces the search range to the range between the smallest and largest section that contain at least one of the searched items. This can be achieved by only comparing with the smallest and largest item of the block. This is repeated until the search range can no longer be reduced. In the second phase, each thread looks for its own item using ordinary binary search, which is initialized with the range determined using $p$-ary search. We call the method \emph{partial} $p$-ary search because only the first iterations of searching are executed in $p$-ary fashion before falling back to standard binary search. Algorithm~\ref{alg:partialParySearch} illustrates the idea.

\begin{algorithm}[caption={Partial $p$-ary search}, label={alg:partialParySearch}, float=tb]
function binarySearch($\langle l_1$, ..., $l_N \rangle$: Ordered list to search in,
            $x$: Item to search, $(a, b)$: Initial search range)
    while $a-b > 1$ do
        $s := (a + b) / 2$
        if $l_s > x$ then $b := s$ else $a := s + 1$
    return a

function partialP-arySearch($\langle l_1$, ..., $l_N \rangle$: Ordered list to search in,
            $\langle x_1$, ..., $x_p \rangle$: Ordered items to search, $t$: Thread index)
    $(a, b) := (0, N)$
    $\langle s_1$, ..., $s_p \rangle$: Pivots of all threads (shared)
    $\langle r_1$, ..., $r_p \rangle$: State of all threads (shared)

    while true do
        $s_t := a + t \cdot (b-a)/(p-1)$
        if $x_0 > l_{s_t}$ then
            $r_t :=$ smaller
        else if $x_p < l_{s_\textrm{t}}$ then
            $r_t :=$ larger
        else
            $r_t :=$ within
        $a := s_m$ where $m$ is the maximum number with $r_m =$ smaller
        $b := s_n$ where $n$ is the minimum number with $r_n =$ larger
        if $n - m$ close to $p$ then break
    return binarySearch($l$, $x_t$, $a$, $b$)
\end{algorithm}

\paragraph{Uncompetitive Method.}
For each of the $s$ threads, the split method searches for the number of heavy items to include. To make use of interleaved addressing, an \emph{inverse} split algorithm would start one thread for each item and check them all in parallel. The method is 60 times slower than the baseline.

\subsection{Pack Method}
\label{sec:pack}
The pack step is similar to sequential alias table construction but starts at a specific position that is determined by the split. As a baseline, we transfer the original pack algorithm of PSA \cite{hubschle2019parallel} to the GPU. We now explain multiple ideas that incrementally improve its performance.

\paragraph{\texttt{l} and \texttt{h} in Shared Memory.}
The baseline pack method accesses the \texttt{l} and \texttt{h} arrays in a way that cannot be coalesced. For the shared memory method, we first copy the array sections that each block will later access to the shared memory in an efficient interleaved fashion. Because shared memory is much faster, the rather inefficient memory access pattern of the pack operation is no longer a problem.

\paragraph{Weight in \texttt{l} and \texttt{h} Arrays.}
In the baseline method, the \texttt{l} and \texttt{h} arrays only store the index of the light and heavy items. The pack method reads items from the arrays and then loads the corresponding weight from the input array. Using the shared memory method above, access to the \texttt{l} and \texttt{h} arrays is cheap but access to the weights array is still expensive and not properly coalesced. Instead of only storing the item index in \texttt{l} and \texttt{h}, we now also store the weight of the items. Because we do this during partitioning into the \texttt{l} and \texttt{h} arrays, no additional passes over the data are required.

\paragraph{Chunked Loading.}
With the shared memory pack method, we assume that the light and heavy items of each section fit into the shared memory. In order to make each section small enough, we need to compute a large number of splits. The idea of the chunked pack method is to generate larger sections and therefore reduce the number of splits required. This can be achieved by efficiently loading chunks of the \texttt{l} and \texttt{h} arrays to shared memory as needed. During packing, whenever all threads of a block have no light or no heavy items left, the threads cooperate to load a new chunk of new data from the global \texttt{l} and \texttt{h} arrays in an interleaved way (see Algorithm \ref{alg:chunkedPack}).

\begin{algorithm}[caption={Chunked pack method}, label={alg:chunkedPack}, float=!tbh]
function chunkedPack()
    while not all threads are finished do
        copyChunks()
        if current thread is not finished then
            packUntilChunkEnd()

function copyChunks()
    foreach worker thread $T$ do
        if $T$ already handled more than $2/3$ of its light items then
            Copy next light items that $T$ will access to shared memory
        if $T$ already handled more than $2/3$ of its heavy items then
            Copy next heavy items that $T$ will access to shared memory

function packUntilChunkEnd()
    $i, j, w$: State like in the PSA method
    Restore state of $i, j, w$
    while true do
        if light or heavy array in shared memory ran out of items then
            Store state of $i, j, w$
            return

        // Normal packing loop, see PSA^\cite{hubschle2019parallel}^
        if $w \leq W/N$ then
            ...
        else
            ...
    Mark thread as finished
\end{algorithm}

\paragraph{Uncompetitive Methods.}
Because the \texttt{l} and \texttt{h} arrays are sorted by item index, write operations to the alias table cannot be coalesced. Writing to the shared memory first and copying the table afterwards is not feasible because split sections can write to overlapping memory locations in the output.\footnote{Without loss of generality, the last processed light item of a thread can have a significantly lower index in the input array than the last processed heavy item. The next thread can then process a light item with an index smaller than the index of the current thread's last heavy item.} Reordering the \texttt{l} and \texttt{h} arrays before executing the split kernel is up to 2.2 times slower than the baseline method. The CPU implementation \cite{hubschle2019parallel} initializes the alias table with the weights instead of accessing the array directly in the pack step. On the GPU, the method is roughly 15\,\% slower than the baseline method. The CPU implementation iterates over the input items to find the next heavy item instead of using the \texttt{l} and \texttt{h} arrays. On the GPU, the method is more than 3.7 times slower than the baseline method. The pack method accesses the weights array indirectly using \texttt{weight[l[i]]} but precomputing those values directly to an array is roughly 10\,\% slower than the baseline method.

\subsection{PSA+}
Hübschle-Schneider's and Sanders' implementation \cite{hubschle2019parallel} executes greedy packing before partitioning into the \texttt{l} and \texttt{h} arrays. For that, it uses the sweeping pack method \cite{hubschle2019parallel}, which is not efficient on GPUs. The idea of our PSA+ implementation is to perform greedy packing while partitioning, when the arrays are already available in the fast shared memory. We then only copy items back to the global \texttt{l} and \texttt{h} arrays that are not yet handled. With this method, we are able to reduce both the time of the prefix sum and the memory reads and writes to the \texttt{l} and \texttt{h} arrays. Our PSA+ implementation does not perform any additional access to global memory that would not have been done with PSA.

\section{Sampling}
We now consider algorithms for efficiently sampling alias tables on the GPU. The baseline sampling method directly follows the algorithm of Walker~\cite{walker1977efficient}, which first chooses a random table row and then either outputs the item or its alias. The throughput scales with the number of samples drawn because table rows that are accessed a second time might already be cached. We now present batched methods that make explicit use of the cache.

\paragraph{Cached Sectioned Sampling.}
To increase the number of cache hits, we use a similar idea as in Algorithm~R \cite{sanders2018efficient}. For uniform sampling, Algorithm R splits the items to be sampled into two sections recursively. The number of samples to be drawn from each section is decided using a binomial deviate. Each thread then only draws samples from one section and therefore accesses more local memory areas. Our new \emph{cached sectioned sampling} algorithm uses the same idea to split the alias table into one section per block. The threads in the block then draw their samples only from that section, relying on the cache to improve sampling throughput. Splitting an alias table is easier than splitting the items themselves because each table row is sampled with the same probability. Like in Algorithm~R, it is possible to determine the sections without communication by using a pseudorandom number generator. The size of the sections serves as a tuning parameter between the number of sections to calculate and the cache hit probability. In our setting ($N \gg 30$), the normal distribution is a good approximation of the binomial distribution \cite{martinez2013accelerating} and computationally much easier to evaluate.

\paragraph{Cached Limited Sectioned Sampling.}
Even if the whole section would theoretically fit into the cache, the cached sectioned sampling method only achieves a small increase in throughput. This is due to multiple blocks being scheduled to each SM and therefore evicting each other's cache entries. Our new cached \emph{limited} sectioned method allocates (but does not use) so much shared memory that only a single block can be executed on each SM. Like the cached sectioned method, the method allows for using the section size as a tuning parameter.

\paragraph{Shared Memory Sectioned Sampling.}
Our shared memory sampling algorithm explicitly copies each block's section to the fast shared memory in an interleaved fashion and then samples from there. The section size is limited by the size of the shared memory, so it cannot be used as a tuning parameter.

\section{Evaluation}
For comparing our methods among each other and with the CPU implementation~\cite{hubschle2019parallel}, we use both consumer devices and powerful servers, as listed in Table~\ref{tab:setup}. For speedups, we compare the \rtx and the high-end desktop CPU because they have a similar price range. Because the behavior is similar on all tested GPUs, we only plot measurements from the \rtx. We use uniform random weights and the shuffled power law distribution ($w_i=i^{-\alpha}$ in random order).

\newlength{\storetabcolsep}
\setlength{\storetabcolsep}{\tabcolsep}
\setlength{\tabcolsep}{3pt}
\begin{table}[t]
    \centering
    \begin{tabular}{ l l }
        \toprule
        Machine      & Hardware specifications \\ \midrule
        Desktop      & AMD Ryzen 3950X (16 cores, 32 threads), Ubuntu 20.04 \\
        AMD server   & AMD EPYC 7551P (32 cores, 64 threads), Ubuntu 20.04 \\
        Intel server & 4x Intel Xeon Gold 6138 (4$\times$20 cores, 160 threads), Ubuntu 20.04 \\ \midrule
        \gtx         & Nvidia GeForce GTX 1650 Super GPU, CUDA 11.1 \\
                     & Intel Xeon 1230 V2 (4 cores), Arch Linux (2021-06-11) \\
        \rtx         & Nvidia GeForce \rtx GPU, CUDA 11.1 \\
                     & Intel Core i5-750 (4 cores), Ubuntu 16.04 \\
        \tesla       & Nvidia \tesla data center GPU, CUDA 11.0 \\
                     & Intel Xeon Gold 6230 (using 4 cores), Red Hat Enterprise 7.7 \\ \bottomrule
    \end{tabular}\\
    \caption{Machines used for the evaluation.}
    \label{tab:setup}
\end{table}
\setlength{\tabcolsep}{\storetabcolsep}

\paragraph{Implementation details.}
By performing only index and pointer arithmetics in conditional branches and accessing the memory afterwards, we achieve a speedup of up to 2. In the pack method, we use casts to \texttt{int4} to help the compiler generate a single 128-bit operation instead of two 64-bit operations for accesses to the alias table rows. This reduces memory transfers by 50\,\% and makes the pack operation nearly 50\,\% faster.

\subsection{Construction}
\begin{figure}[t]
    \centering
    \includegraphics[width=0.8\textwidth]{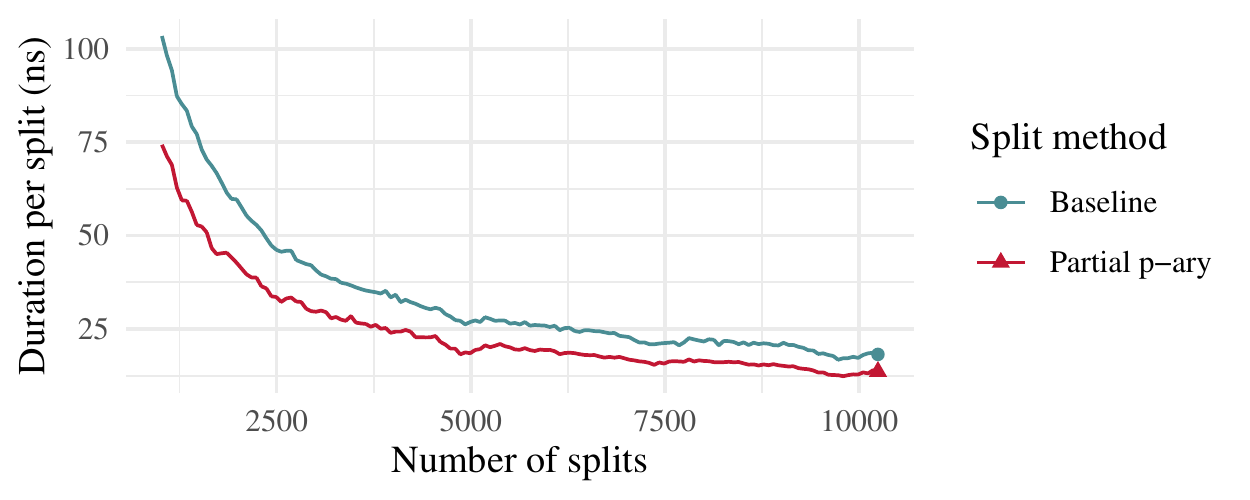}
    \caption[Time for splitting]{Time needed for determining a single split using different split algorithms. Using $10^7$ input items with uniform random weights.}
    \label{fig:buildSpeedSplitVariants}
\end{figure}

\begin{figure}[t]
    \centering
    \includegraphics[width=0.85\textwidth]{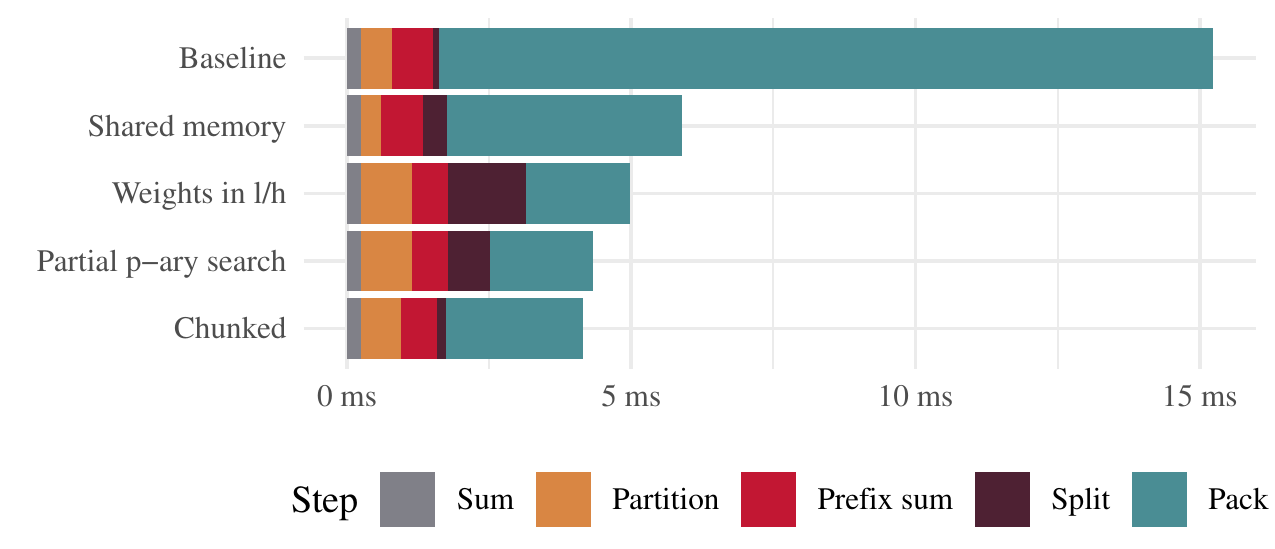}
    \caption[Construction duration overall comparison]{Construction duration for a table of size $10^7$ with uniform random weights.}
    \label{fig:speedBuildHistoric}
\end{figure}

A comparison of our split methods is plotted in Figure \ref{fig:buildSpeedSplitVariants}. Independently of the number of splits $s$, the partial $p$-ary split method is up to 1.5 times faster than the baseline method, depending on the input distribution and number of splits. Figure~\ref{fig:speedBuildHistoric} shows how the techniques of Section \ref{sec:pack} achieve a speedup of 3.7 to the baseline. Because the pack method has an influence on the number of splits to calculate, the figure shows the full construction time including splitting. When storing weights in \texttt{l} and \texttt{h}, the pack step gets 2 times faster while the split and partition steps get 2 times slower because of an increased size of array elements. In total, this results in a speed improvement because the pack step takes most time overall. The pack step of the chunked method is slower than the shared memory method because its memory access cannot be coalesced as well but it speeds up the splitting step significantly. For large $N$, the chunked method is slightly faster than the shared memory method.

\paragraph{PSA+.}
When the items have uniform random weights, PSA+ on GPUs can greedily handle around 90\,\% of the items. A reason why Hübschle-Schneider and Sanders' \cite{hubschle2019parallel} algorithm can pack a higher fraction of the items greedily is that our section size is limited by the shared memory and therefore rather small. We only attempt greedy packing in promising situations by introducing a threshold for the minimum number of light and heavy items in each section. Using uniform random weights with $10^7$ items, PSA+ achieves a speedup of 1.5 to PSA and using a shuffled power law distribution with exponent $\alpha=0.5$, it achieves a speedup of 1.4. While PSA+ can be slower for some weight distributions, it achieves significant speedups for these important distributions.

\paragraph{Comparison with the CPU method.}
Our GPU-based chunked method achieves a speedup of 17 on the \rtx over Ref. \cite{hubschle2019parallel} on a desktop CPU, as listed in Table~\ref{tab:speedConstructionVsLHS}. Constructing with $N > 10^6$ items, our method is faster even when including the time to transfer the input weights to the GPU. In fact, our construction is faster than the time needed to transfer a finished alias table to the GPU.

\begin{table}[t]
    \centering
    \begin{tabular}{ l r r }
        \toprule
        Machine      & \multicolumn{2}{c}{Construction time} \\
        ~            & \cc{$N=10^7$} & \cc{$N=10^8$} \\ \midrule
        Desktop CPU  &       69.2 ms &      743.2 ms \\
        AMD server   &       21.3 ms &      151.5 ms \\
        Intel server &       18.2 ms &       83.1 ms \\ \midrule
        \gtx         &        7.6 ms &            --\footnotemark \\
        \rtx         &        4.0 ms &       32.8 ms \\
        \tesla       &        2.5 ms &       23.9 ms \\ \bottomrule
    \end{tabular}
    \caption[Construction duration on different machines]{Construction duration comparison with the CPU method \cite{hubschle2019parallel}. Input are $10^7$ and $10^8$ items with a shuffled power law distributed weights.}
   \label{tab:speedConstructionVsLHS}
\end{table}
\footnotetext{Not enough memory for temporary data structures during construction.}

\subsection{Sampling}
\begin{figure}[!tp]
    \centering
    \subfigure[$N=10^6$ items]{\includegraphics[width=0.48\textwidth]{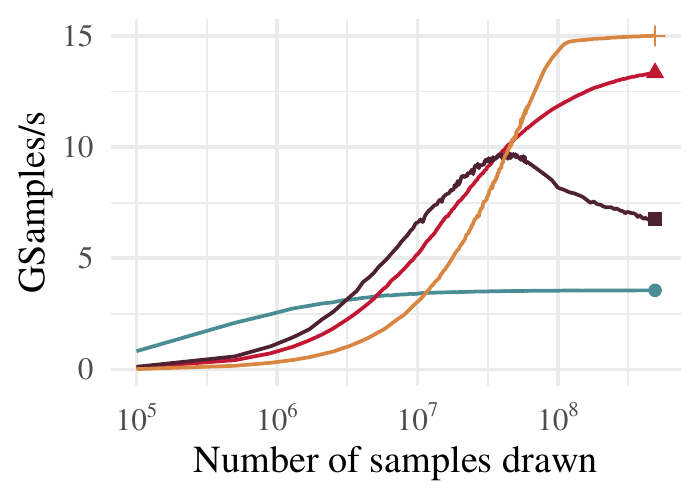}}
    \subfigure[$N=10^7$ items]{\includegraphics[width=0.48\textwidth]{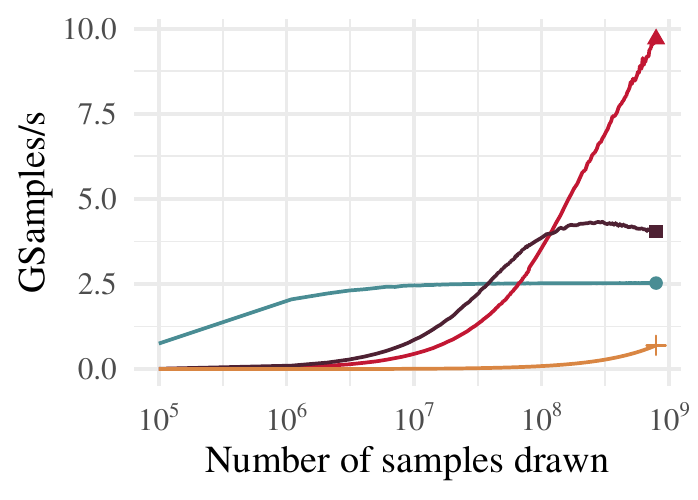}}
    \includegraphics[width=\textwidth]{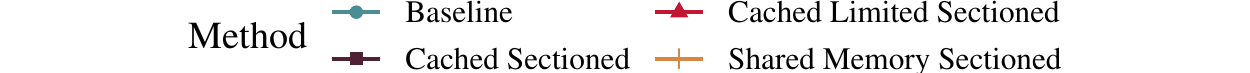}
    \caption[Sampling throughput]{Comparison between sampling methods depending on the input size and number of samples drawn. Input is a uniform random weight distribution. Note the logarithmic x-axes.}
    \label{fig:samplesPerSecondSamplerMethod}
\end{figure}

\begin{figure}[!tp]
    \centering
    \includegraphics[width=0.95\textwidth]{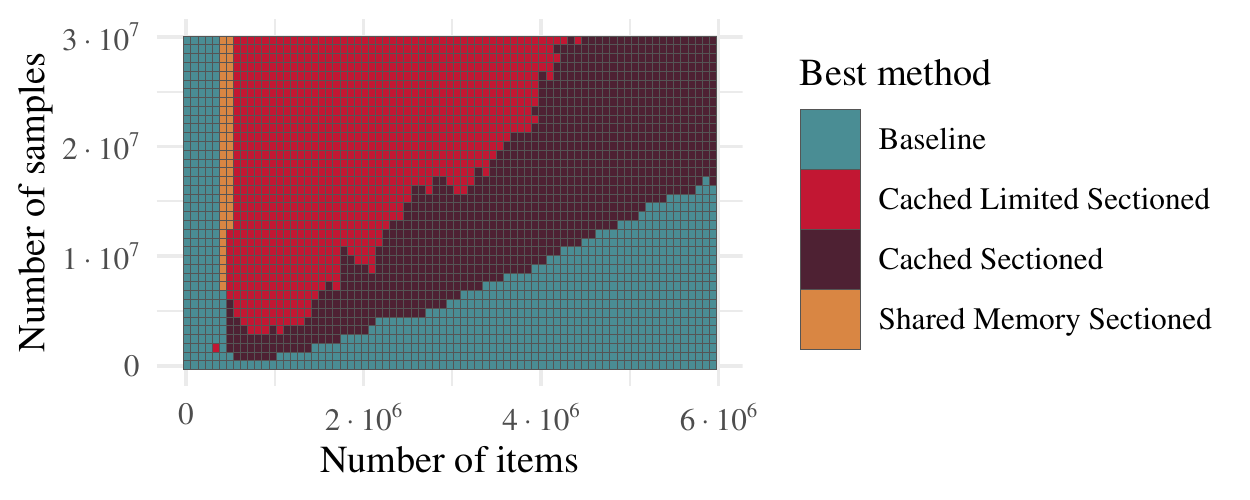}
    \caption[Heatmap of sampling method with highest sampling throughput]{Comparison which method has the highest throughput depending on table size and number of samples drawn. The input weights are drawn from a uniform random distribution.}
    \label{fig:samplesPerSecondMethodHeatmap}
\end{figure}

Figure \ref{fig:samplesPerSecondSamplerMethod} shows a comparison of the baseline sampling method and the three sectioned methods. The baseline method does not need preprocessing and is therefore fastest for small numbers of samples. The sectioned methods have significant startup overhead for determining the sections or copying data but if the number of samples drawn is increased, the investment pays off. Figure \ref{fig:samplesPerSecondMethodHeatmap} shows the best method for varying table size and number of samples. While the shared memory sectioned method can achieve higher peak throughputs, the cached limited sectioned method is more generic and achieves a good throughput in more cases.

\paragraph{Comparison with the CPU method.}
Table~\ref{tab:speedSamplingVsLHS} compares the throughput of our sectioned limited method with the CPU implementation of Ref. \cite{hubschle2019parallel} when using shuffled power law distributed weights. Our GPU method has up to 24 times more throughput on the \rtx than Ref. \cite{hubschle2019parallel} on the desktop CPU. Even for large $N$, we can outperform the expensive Intel server using consumer hardware.

\begin{table}[t]
    \centering
    \begin{tabular}{ l r r r r }
        \toprule
        Machine  & \multicolumn{4}{c}{GSamples/s} \\
                 & \cc{$N=10^6$} & \cc{$N=10^7$} & \cc{$N=10^8$} & \cc{$N=10^9$} \\ \midrule
        Desktop CPU  &       3.67 &       0.42 &       0.37 &       0.37 \\
        AMD server   &       1.36 &       0.92 &       0.92 &       0.89 \\
        Intel server &       7.98 &       2.67 &       2.17 &       1.63 \\ \midrule
        \gtx         &       6.41 &       3.18 &       1.06 &         -- \\
        \rtx         &      13.43 &      10.14 &       2.44 &         -- \\
        \tesla       &     106.71\footnotemark\hspace{-1.7mm}
                                  &      26.93 &       5.62 &         -- \\ \bottomrule
    \end{tabular}\\
    \caption[Sampling throughput on different machines]{Sampling throughput comparison with the CPU method. Drawing $10^9$ samples from a table of varying size. On the GPU, we use our fastest variant for each input size ($N \leq 10^7$: cached limited sectioned, $N=10^8$: baseline).}
    \label{tab:speedSamplingVsLHS}
\end{table}
\footnotetext{Throughput with N=$10^6$ is only constrained by the 64-bit floating point unit, which is significantly faster on the \tesla than on the other cards.}

\subsection{Power Consumption}
\begin{table}[t]
    \centering
    \begin{tabular}{ l r r }
        \toprule
        Machine      & Construction       & Sampling      \\ \midrule
        Desktop CPU  &  98~J/$10^8$ items & 376~J/GSample \\
        AMD server   &  25~J/$10^8$ items & 181~J/GSample \\
        Intel server &  45~J/$10^8$ items & 242~J/GSample \\ \midrule
        \gtx         & $\approx$ 7~J/$10^8$ items\footnotemark\hspace{-1.7mm}
                                          &  92~J/GSample \\
        \rtx         &   7~J/$10^8$ items &  69~J/GSample \\
        \tesla       &   5~J/$10^8$ items &  28~J/GSample \\ \bottomrule
    \end{tabular}\\
    \caption{Power usage of constructing an alias table of size $10^8$ with shuffled power law distributed weights and drawing $10^9$ samples}
    \label{tab:powerUsage}
\end{table}
\footnotetext{Not enough memory for temporary data structures during construction. Extrapolation based on a measurement with $N=6 \cdot 10^7$.}

Because of their different architecture, comparing only running time between GPUs and CPUs can be unfair. A good sanity check is to compare by energy consumption, which is independent of current market prices and covers a major cost factor of computing. To compensate for different hardware setups, we calculate the CPU power usage by the difference between idle and loaded state using external measurements. For the GPUs, we directly use the values reported by the cards, adding additional 40~W to account for the CPUs that manage the cards.\footnote{Based on external measurements with the \rtx.} Table~\ref{tab:powerUsage} lists the power usage measurements of construction and sampling.

\section{Conclusions}
In this report, we have presented new algorithms that make construction of and sampling from alias tables efficient on GPUs. We are able to achieve a speedup of 17 to the CPU implementation of Hübschle-Schneider and Sanders \cite{hubschle2019parallel}, while simultaneously being more energy-efficient. We introduce a new search algorithm, partial $p$-ary search, that enables fast splitting. Our pack method with chunked loading to the shared memory adapts the memory access pattern to be more efficient on GPUs. Our sectioned limited sampling algorithm is up to 24 times faster than the CPU implementation. This is achieved by dividing the alias table into sections which can then be sampled in a more cache-efficient way. In the future, we plan to evaluate our methods in real-world applications such as graph generation and also evaluate partial $p$-ary search on its own.

\section{Acknowledgments}
The authors acknowledge support by the state of Baden-Württemberg through bwHPC.
This project has received funding from the European Research Council (ERC) under the European Union’s Horizon 2020 research and innovation programme (grant agreement No. 882500).
We also thank Emanuel Schrade for co-supervising the thesis that this report is based on.

\includegraphics[width=4cm]{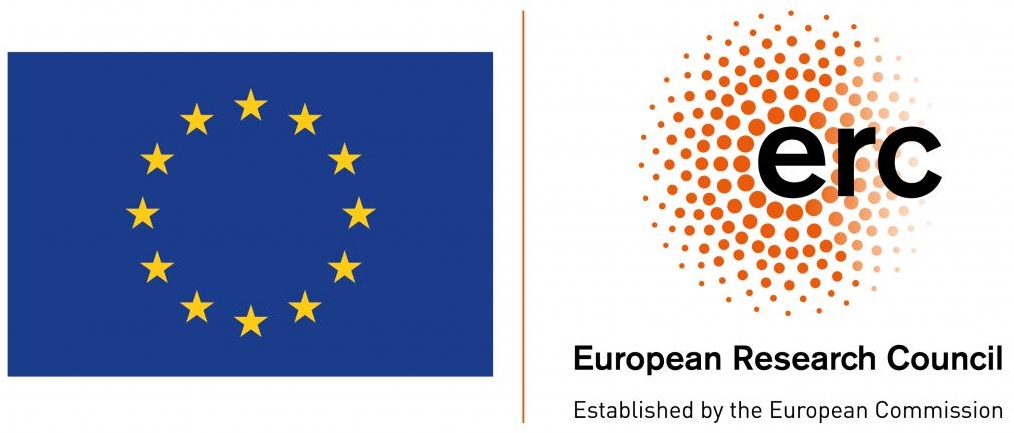}

\bibliography{paper}

\begin{thebibliography}{10}

\bibitem{nvidia2018turingArchitecture}
{Nvidia} {Turing} {GPU} architecture.
\newblock
  \url{https://images.nvidia.com/aem-dam/Solutions/design-visualization/technologies/turing-architecture/NVIDIA-Turing-Architecture-Whitepaper.pdf},
  2018.
\newblock Accessed: 2020-12-14.

\bibitem{nvidia2020bestpractices}
{CUDA} {C++} best practices guide.
\newblock
  \url{https://docs.nvidia.com/cuda/pdf/CUDA_C_Best_Practices_Guide.pdf}, 2020.
\newblock Accessed: 2020-07-15.

\bibitem{binder2019massively}
Nikolaus Binder and Alexander Keller.
\newblock Massively parallel construction of radix tree forests for the
  efficient sampling of discrete probability distributions.
\newblock {\em arXiv preprint arXiv:1901.05423}, 2019.

\bibitem{burke2004bidirectional}
David Burke, Abhijeet Ghosh, and Wolfgang Heidrich.
\newblock Bidirectional importance sampling for illumination from environment
  maps.
\newblock In {\em ACM SIGGRAPH 2004 Sketches}, page 112. 2004.

\bibitem{galerne2012gabor}
Bruno Galerne, Ares Lagae, Sylvain Lefebvre, and George Drettakis.
\newblock Gabor noise by example.
\newblock {\em ACM Transactions on Graphics (TOG)}, 31(4):1--9, 2012.

\bibitem{hubschle2019parallel}
Lorenz H{\"{u}}bschle{-}Schneider and Peter Sanders.
\newblock Parallel weighted random sampling.
\newblock In {\em 27th Annual European Symposium on Algorithms, {ESA}}, volume
  144 of {\em LIPIcs}, pages 59:1--59:24. Schloss Dagstuhl - Leibniz-Zentrum
  f{\"{u}}r Informatik, 2019.

\bibitem{hubschle2019linear}
Lorenz H{\"{u}}bschle{-}Schneider and Peter Sanders.
\newblock Linear work generation of {R-MAT} graphs.
\newblock {\em Netw. Sci.}, 8(4):543--550, 2020.

\bibitem{kaldewey2009parallel}
Tim Kaldewey, Jeff Hagen, Andrea Di~Blas, and Eric Sedlar.
\newblock Parallel search on video cards.
\newblock In {\em First USENIX Workshop on Hot Topics in Parallelism
  (HotPar'09)}, 2009.

\bibitem{sourceGitHub}
Hans-Peter Lehmann.
\newblock {ByteHamster} / alias-table-gpu.
\newblock \url{https://github.com/ByteHamster/alias-table-gpu}, 2021.

\bibitem{lehmann2021alias}
Hans-Peter Lehmann.
\newblock Weighted random sampling - alias tables on the {GPU}.
\newblock Master's thesis, Karlsruher Institut für Technologie (KIT), 2021.

\bibitem{li2017saberlda}
Kaiwei Li, Jianfei Chen, Wenguang Chen, and Jun Zhu.
\newblock {SaberLDA}: Sparsity-aware learning of topic models on {GPUs}.
\newblock {\em ACM SIGPLAN Notices}, 52(4):497--509, 2017.

\bibitem{martinez2013accelerating}
MA~Mart{\'\i}nez-del Amor.
\newblock {\em Accelerating membrane systems simulators using high performance
  computing with GPU}.
\newblock PhD thesis, University of Seville, 2013.

\bibitem{mohanty2012efficient}
Siddhant Mohanty, AK~Mohanty, and F~Carminati.
\newblock Efficient pseudo-random number generation for monte-carlo simulations
  using graphic processors.
\newblock In {\em Journal of Physics: Conference Series}, volume 368, page
  012024. IOP Publishing, 2012.

\bibitem{nguyen2007gpu}
Hubert Nguyen.
\newblock {\em {GPU} gems 3}.
\newblock Addison-Wesley Professional, 2007.

\bibitem{nvidia2008gettingstarted}
Greg Ruetsch and Brent Oster.
\newblock Getting started with cuda.
\newblock
  \url{https://www.nvidia.com/content/cudazone/download/Getting_Started_w_CUDA_Training_NVISION08.pdf},
  2008.
\newblock Accessed: 2020-07-15.

\bibitem{sanders2018efficient}
Peter Sanders, Sebastian Lamm, Lorenz H{\"u}bschle-Schneider, Emanuel Schrade,
  and Carsten Dachsbacher.
\newblock Efficient parallel random sampling—vectorized, cache-efficient, and
  online.
\newblock {\em ACM Transactions on Mathematical Software (TOMS)}, 44(3):1--14,
  2018.

\bibitem{vose1991linear}
Michael~D. Vose.
\newblock A linear algorithm for generating random numbers with a given
  distribution.
\newblock {\em IEEE Transactions on Software Engineering}, (9):972--975, 1991.

\bibitem{walker1977efficient}
Alastair~J Walker.
\newblock An efficient method for generating discrete random variables with
  general distributions.
\newblock {\em ACM Transactions on Mathematical Software (TOMS)},
  3(3):253--256, 1977.

\bibitem{wilderman2007method}
SJ~Wilderman and YK~Dewaraja.
\newblock Method for fast {CT}/{SPECT}-based {3D} {M}onte {C}arlo absorbed dose
  computations in internal emitter therapy.
\newblock {\em IEEE Transactions on Nuclear Science}, 54(1):146--151, 2007.

\end{thebibliography}
\end{document}